\documentclass[letterpaper]{article} 
\usepackage{aaai2026}  
\usepackage{times}  
\usepackage{helvet}  
\usepackage{courier}  
\usepackage[hyphens]{url}  
\usepackage{graphicx} 
\usepackage{natbib}  
\usepackage{caption} 
\usepackage{algorithm}
\usepackage{algorithmic}

\usepackage{newfloat}
\usepackage{listings}
\DeclareCaptionStyle{ruled}{labelfont=normalfont,labelsep=colon,strut=off} 
\floatstyle{ruled}
\newfloat{listing}{tb}{lst}{}
\floatname{listing}{Listing}
\usepackage{subcaption}
\usepackage{tikz}
\usepackage{pgfplots}
\usepackage{csquotes}
\usepackage{multirow}
\usepackage{booktabs}
\usepackage{makecell}
\newcommand{\codecount}[1]{#1}
\usetikzlibrary{shapes, arrows.meta, positioning}
\usetikzlibrary{pgfplots.groupplots}
\definecolor{plot_color_1}{RGB}{230,97,1}
\definecolor{plot_color_2}{RGB}{253,184,99}
\definecolor{plot_color_3}{RGB}{178,171,210}
\definecolor{plot_color_4}{RGB}{94,60,153}
\usepackage{xurl}
\title{Hands‑Off or Hands‑On? Variation in Area Chair Practices and Implications for AI Support}
\author {
    Ines Arous\textsuperscript{\rm 1},
    Neha Nayak Kennard\textsuperscript{\rm 2},
    Andrei Mircea\textsuperscript{\rm 3},
    Emily Kuang\textsuperscript{\rm 1},
    Jackie Chi Kit Cheung\textsuperscript{\rm 4},
    Andrew McCallum\textsuperscript{\rm 2}
}
\affiliations {
    \textsuperscript{\rm 1}York University\\
    \textsuperscript{\rm 2}University of Massachusetts Amherst\\
    \textsuperscript{\rm 3}University of Montreal, Mila\\
    \textsuperscript{\rm 4}McGill University, Canada CIFAR AI Chair, Mila\\
    inesar@yorku.ca, kennard@cs.umass.edu, mirceara@mila.quebec,  ekuang@yorku.ca, cheungja@mila.quebec,
    mccallum@cs.umass.edu
}

\begin{document}

\maketitle

\begin{abstract}
Area chairs (ACs) play a critical role in the peer-review process, managing conflicts and ensuring fair outcomes. Although AI tools have been proposed to support ACs, little is known about the challenges they face and their perceptions of these technologies. In this paper, we conduct interviews including a design probe with 27 ACs in AI to explore their challenges, strategies, and perspectives on potential AI tools. Through thematic analysis, we identify key tensions arising from the growing volume of submissions, uneven reviewer expertise, and the complex task of managing the relationship between reviewers and authors. Most importantly, we find substantial variation in how ACs engage with submissions and influence outcomes: some adopt a largely hands-off approach, while others take a more hands-on role in guiding discussions and decisions. This variation challenges the notion of a single, universal AC practice and highlights the need to account for diverse approaches. When reflecting on the potential use of AI tools, ACs expressed a cautious stance, drawing on their domain knowledge and heightened awareness of AI limitations. From these findings, we derive three design implications: tailoring AI assistance to diverse AC practices, design assistance for discussion moderation, and embedding human‑centered AI principles that preserve human agency in decision‑making.
\end{abstract}

\section{Introduction}
Scientific peer review is pivotal to maintaining quality standards for academic publication. Within this process, area Chairs (ACs) play a critical role: they mediate conflicts, synthesize divergent reviews, and ensure a fair decision for each submission~\cite{ailamaki_sigmod_2019,resnik_ensuring_2016}. All of these responsibilities are carried out under tight deadlines and an ever-growing number of submissions, making their role both demanding and time-sensitive~\cite{galipeau_scoping_2016, tran_open_2020}. For instance, the International Conference on Learning Representations (ICLR) 2025 received 11,603 submissions~\cite{miley}, which is an exponential increase from just 490 submissions in 2017.
At AAAI, submissions reached 29,000 in 2025 compared with just 1,991 in 2015~\cite{AAAI26}. This unprecedented scale places compounding pressures on the review system, from limited workflow support and reviewer scarcity to the difficulty ACs face in maintaining fairness under time pressure.

Various tools have been designed to support the peer review system by providing scaffolding for \textit{reviewers} to enhance the quality of their feedback~\cite{hicks_framing_2016,ngoon_interactive_2018,sun_reviewflow_2024} and to support \textit{authors} in interpreting peer reviews~\cite{yang_understanding_2025,yen_decipher_2020,yen_listen_2017}. Comparatively little attention has been paid to designing tools for the \textit{distinctive needs of ACs}, whose role requires making high‑stakes decisions under tight time constraints. At the same time, substantial work has explored the use of AI, and more recently, large language models (LLMs), to support or mimic components of the AC decision-making process, such as predicting acceptance decisions~\cite {checco_ai-assisted_2021} and summarizing reviews~\cite{li_towards_2023,li_sentiment_2024,zeng_scientific_2025}. Some of these techniques have been deployed in a recent pilot program in AAAI-26~\cite{ellison_aaai_2025}.
While these approaches promise efficiency gains, they often overlook existing AC practices and willingness to adopt such tools, as well as the risks posed by poor integration, including the reinforcement of existing biases~\cite{kuznetsov_what_2024,fecher_friend_2025}.



Our work seeks to understand ACs’ challenges, strategies, and expectations, and explore how potential designs for AI assistance might align or conflict with their real-world practices.  Unlike prior work that proposed automated methods without engaging ACs, this study centers their perspectives to inform the design of future AI assistants. Specifically, our study was guided by the following research questions (RQs):
\begin{itemize}
    \item RQ1: How do ACs experience and navigate their role?
    \begin{itemize}
        \item RQ1a: What challenges do ACs currently face? 
        \item RQ1b: What are their current strategies for addressing these challenges?
    \end{itemize}
    \item RQ2: What are ACs' perceptions of AI tools, and what ethical concerns do they associate with their use?
\end{itemize}

We conducted 27 semi‑structured interviews with a design probe with ACs from ICLR, a uniquely suitable context for this study. ICLR is the only major AI venue that publicly releases reviews and AC feedback for both accepted and rejected papers, making it a frequent testbed for AI‑based peer‑review tools~\cite{arous_peer_2021,zeng_scientific_2025}. 
Through these interviews, we observed substantial variation in the extent of AC engagement with submissions: some ACs took a hands-on approach and actively advocated for promising work, whereas others were hands-off, adopting a judge-like stance and deferring to reviewer consensus. We also found that nudging reviewers was widely seen as the most challenging task, described as frustrating and misaligned with ACs’ seniority.
We further employed a design probe using wireframes of four potential AI‑assistive tools to elicit ACs’ design considerations. ACs expressed cautious optimism about AI assistance, recognizing its potential to alleviate cognitive burden while voicing concerns about bias amplification, misinterpretation, and over-reliance without appropriate safeguards. 
They further emphasized the importance of personalizing AI support, as there is no one-size-fits-all solution for assisting ACs. 
Drawing on these perspectives, we distill key design implications for future AI assistants that support, rather than undermine, AC judgment.


\section{Related Work}
\subsection{AI methods for Scientific Peer Review}
Various tools have been designed to support the peer review process by assisting reviewers and authors. For instance, some methods have explored introducing interactive techniques such as reusable suggestions and adaptive guidance~\cite{ngoon_interactive_2018}, or scaffolding novice reviewers through contextual reflections~\cite{sun_reviewflow_2024}. Other efforts have focused on supporting authors in interpreting and responding to peer reviews, by generating a positive summary of the reviews~\cite{yang_understanding_2025} or by assisting authors in identifying priorities and reflecting on their work~\cite{yen_decipher_2020,yen_listen_2017}.

While various tools have been developed to assist both reviewers and authors, few design studies have addressed the \emph{unique and often conflicting priorities that ACs must navigate when balancing the needs of both groups}. Recent work has begun to explore AI-driven approaches to assist ACs. One such approach involves calibrating reviewers’ scores~\cite{arous_peer_2021,flach_novel_2010,tan_least_2021}, which helps ACs assess the reliability of each review. Other approaches analyze review and rebuttal discourse structure to provide ACs with an overview of the main arguments~\cite{wang_sentiment_2018,ghosal_deepsentipeer_2019,darrin_glimpse_2024, kennard-etal-2022-disapere}. More recently, various efforts have been dedicated to mimicking the ACs' role by predicting the acceptance decisions of submissions~\cite {checco_ai-assisted_2021} and summarizing the reviews with LLMs~\cite{li_towards_2023,li_sentiment_2024,zeng_scientific_2025, sun_metawriter_2024}. Some of these methods have been integrated into the pilot AI Review program at AAAI-26, demonstrating AI's potential to make meaningful contributions to peer review~\cite{biswas_ai-assisted_2026}.

 However, prior efforts have largely relied on speculative assumptions about ACs’ challenges rather than insights from design research or direct engagement with ACs. As a result, these approaches often fail to reflect the nuanced realities of AC work. To address this gap, our study engages directly with ACs through interviews and a design probe, uncovering their challenges and collaboratively identifying opportunities for AI-driven design to support their roles.

\subsection{Human-AI Collaboration for Decision Making in High-Stakes Domains}
Many human-AI collaboration tools have been designed to assist decision-makers in high-stakes domains, including education, medicine, and law~\cite{fok_scim_2023,norkute_towards_2021,fogliato_who_2022,grgic-hlaca_human_2019}. Some studies have demonstrated that effective collaboration between decision-makers and AI can improve decision-making processes under time constraints~\cite {jacobs_designing_2021, bansal_does_2021, kawakami_why_2022}. These tools can provide invaluable support by highlighting relevant information~\cite{yang_harnessing_2023,burgess_healthcare_2023,han_ascleai_2024} and effectively mitigating potential biases~\cite{zhang_designing_2015, wang_factors_2020}. However, other studies show that human-AI decision-making may fail to improve performance or even lead to incorrect decisions in high-stakes domains compared with either human or AI decisions alone~\cite{cheng_how_2022,green_disparate_2019,kawakami_improving_2022,green_principles_2019, grgic-hlaca_human_2019}. In several studies, decision-makers’ attitudes toward AI contribute to shaping the outcome: excessive skepticism can lead to under-reliance on useful predictions, while overconfidence can result in over-reliance on erroneous recommendations~\cite{cabitza_ai_2023,dietvorst_algorithm_2015,kawakami_why_2022}.

To mitigate these risks, prior research has explored design strategies that empower decision-makers to critically evaluate and, when necessary, override AI predictions. These designs aim to preserve human agency by presenting AI-generated recommendations while leaving the final judgment to humans~\cite{chandrasekharan_crossmod_2019, grgic-hlaca_human_2019}. Other approaches seek to mitigate the risk of decision‑makers’ over‑reliance on these systems through a range of design strategies. These include timing AI assistance~\cite{fogliato_who_2022}, encouraging active participation of decision-makers~\cite{solomon_customization_2014}, and seeking support from external sources, such as color-coded tags~\cite{bansal_does_2021,fok_scim_2023,fogliato_who_2022}, or presenting additional context relevant to the decision~\cite{head_augmenting_2021,cai_hello_2019,yang_harnessing_2023}. These strategies help users quickly identify key information without searching or reading entire documents. An alternative approach is to provide concise summaries of relevant information~\cite{norkute_towards_2021,zhang_concepteva_2023}, which help reduce cognitive load and improve decision-making efficiency~\cite{han_ascleai_2024}. Other studies highlight the importance of personalizing AI assistance to decision-makers’ workflows and needs~\cite{corti_it_2024, bertrand_selective_2023, tankelevitch_metacognitive_2024}. For instance, \citet{tankelevitch_metacognitive_2024} argues that personalization in generative AI systems supports users’ metacognitive processes by enabling them to adapt outputs to their workflows while calibrating their confidence in the system. Similar findings have been reported in clinical settings~\cite{corti_it_2024, choi_human-centered_2026}, where personalization mechanisms help preserve human agency by aligning AI systems with local practices and different stages of decision-making workflows. Despite these efforts, it remains an open question how to effectively foster human-AI decision-making, particularly in high-stakes, multi-party contexts like peer review. AC decision‑making is distinctive in that it requires balancing domain expertise with deference to reviewer consensus, all while preserving reviewer anonymity and operating under tight deadlines. 



\section{Method}
We conducted semi-structured interviews with 27 ACs from machine learning conferences. To elicit ACs’ intuitions and perceptions about AI support, the interviews incorporated a design probe using wireframes of four potential AI-assistive tools. This approach allowed participants to envision how AI could fit into their workflows without the constraints of fully implemented prototypes. The speculative nature of the wireframes enabled exploration of broader challenges and perspectives beyond any single design. Each wireframe depicted different modes of AI use, based on existing methods across various reviewing sub-tasks. This research project was reviewed and approved by our institution's IRB.

 \begin{figure}
    \centering
    \includegraphics[scale=1.38]{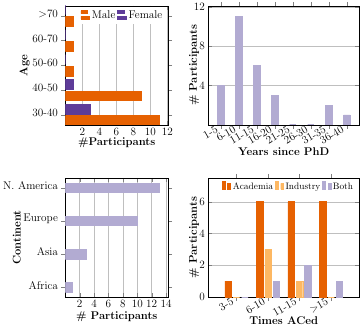}
    \caption{Demographics of the ACs we interviewed. Our participants represented diverse ages, genders, affiliations, and seniority levels.}
    \label{fig:demographics}
\end{figure}
\subsection{Participants}
Our intended participant pool was researchers who had previously served as ACs for machine learning conferences. To reach this specific group, we contacted ACs for ICLR, a top-tier machine learning conference (see Sect.~\ref{sec:iclr_process} for details on the ICLR reviewing process).
Although the inclusion criterion was having previously served as an AC for ICLR, it is worth noting that all participants had served as ACs for multiple machine learning conferences.
During the interviews, we encouraged them to draw on their experiences across these various conferences. 
Using ICLR as an inclusion criterion enabled us to leverage the publicly available lists of ACs on ICLR websites\footnote{e.g.,\url{https://iclr.cc/Conferences/2022/ProgramCommittee}} to retrieve a list of 561 former ACs. Since ICLR is hosted on OpenReview, we also used the OpenReview API\footnote{\url{https://docs.openreview.net/}} to retrieve the homepages of these researchers. We manually collected publicly available email addresses of as many researchers on this list as possible, found either on their homepages or recent publications. 

This process resulted in a collection of 540 email addresses. Expecting a very low response rate, we contacted all 540 candidates by email, explaining the goal of our study and requesting their participation in a semi-structured interview. In total, 58 potential participants responded to our email, of whom 28 consented to participate in the study. The demographics of all participants are depicted in Figure~\ref{fig:demographics}. We list aggregated characteristics of study participants to prevent deanonymization. All participants reported their gender identity as either \textit{man} or \textit{woman}, and their ages ranged from 30 to 70 years or older. They were based across four continents: Africa, Asia, Europe, and North America. All had served as ACs at least three times. Nineteen participants were affiliated with academia, four with industry, and four held joint appointments in both sectors.
The interviews were conducted in August and September 2023.

\begin{figure*}
        \centering
        \begin{subfigure}[b]{0.45\textwidth}
            \centering
            \includegraphics[width=\textwidth]{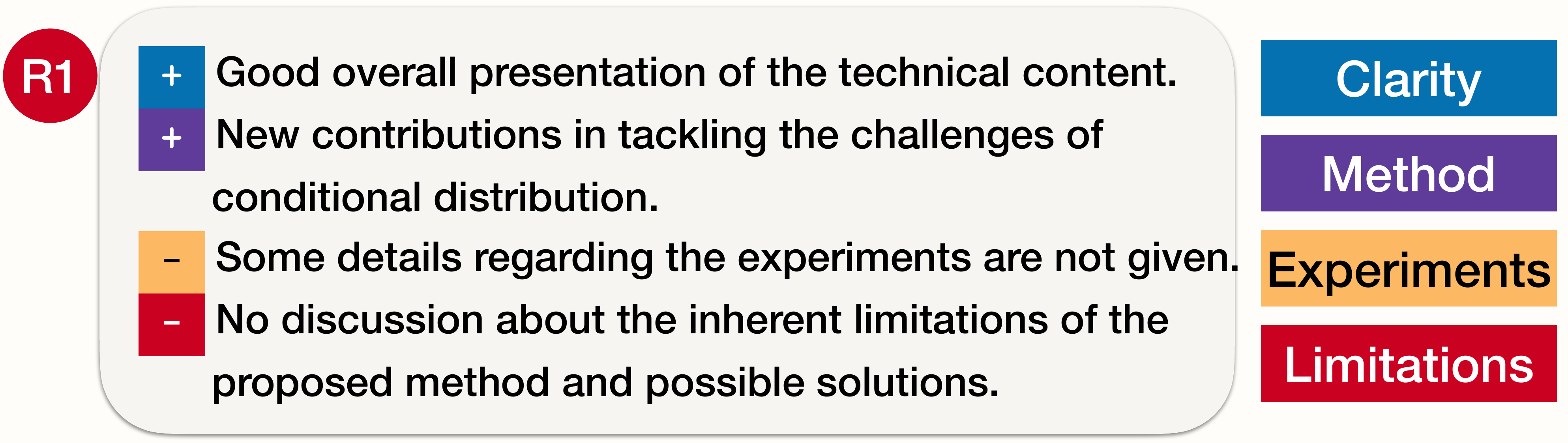}
            \caption{Argument Labels}  
            \label{fig:argument}
        \end{subfigure}
        \begin{subfigure}[b]{0.45\textwidth}  
            \centering
            \includegraphics[width=\textwidth]{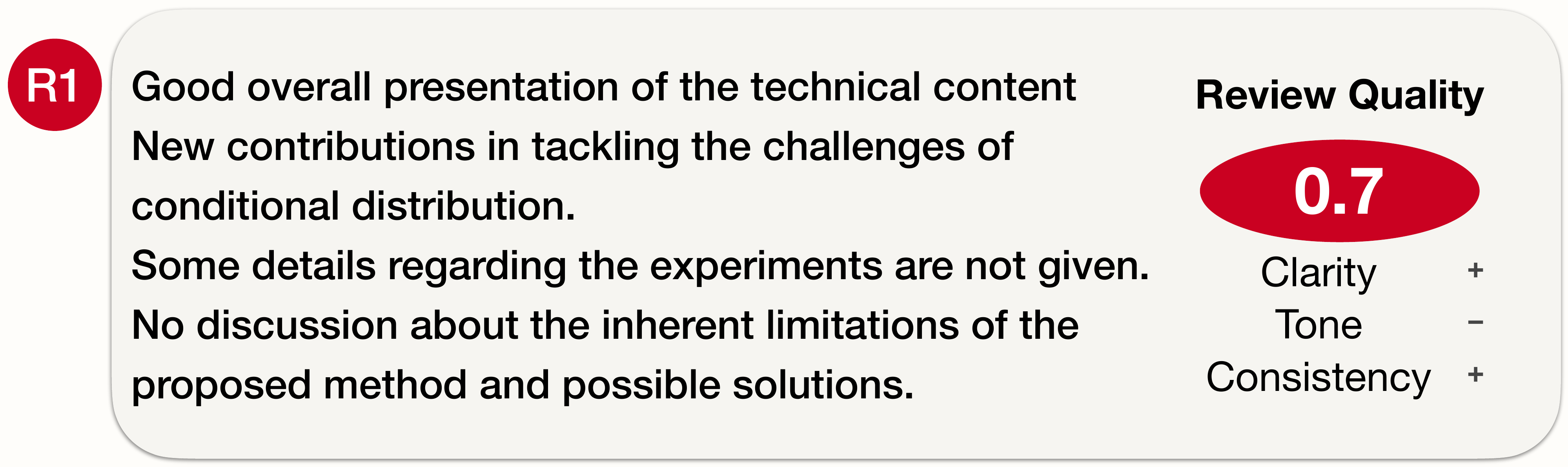}
            \caption{Conformity to Standards}
            \label{fig:standards}
        \end{subfigure}
        \vskip\baselineskip
        \begin{subfigure}[b]{0.45\textwidth}   
            \centering 
            \includegraphics[width=\textwidth]{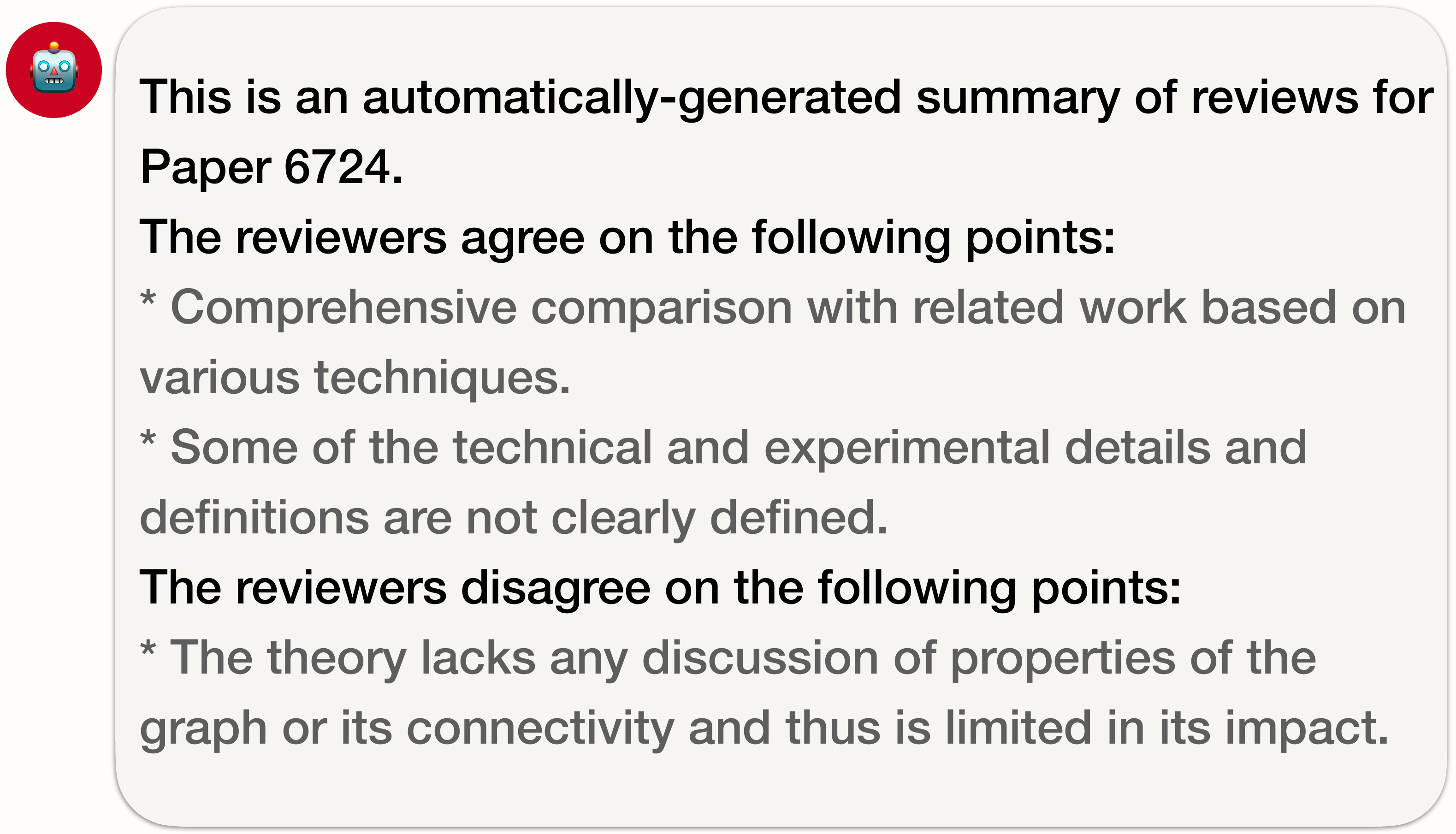}
            \caption{Generated Summary}
            \label{fig:generated}
        \end{subfigure}
        \begin{subfigure}[b]{0.45\textwidth}   
            \centering 
            \includegraphics[width=\textwidth]{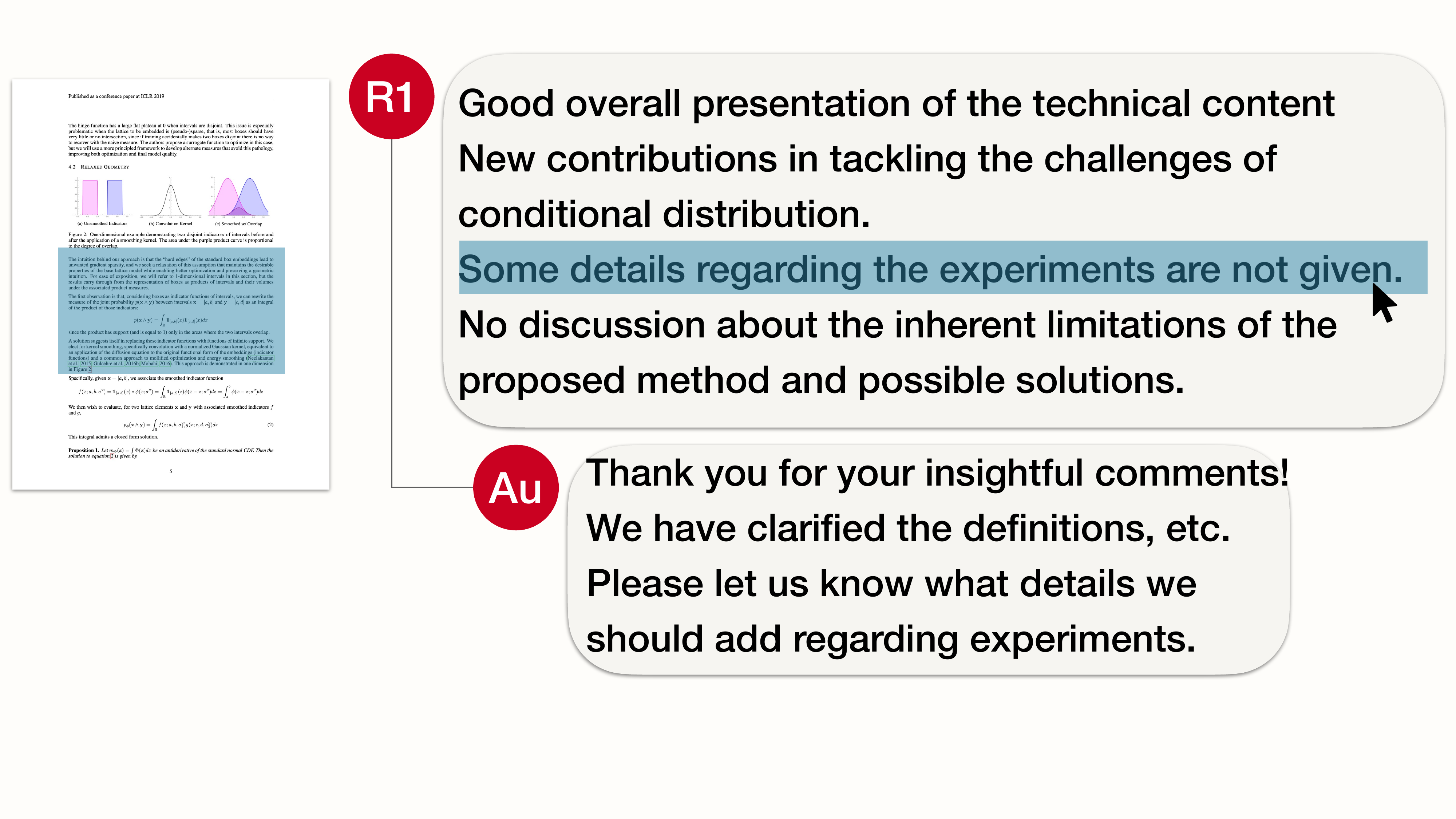}
            \caption{Comment Context}   
            \label{fig:context}
        \end{subfigure}
        \caption{The wireframes shown to participants: (a) Argument labels: each argument within a review is associated with a color indicating the discussed aspect (e.g., clarity, method), and is accompanied by a (+) or (-) sign to denote its sentiment, (b) Conformity to standards: each review is associated with a number reflecting its conformity to a conference standards, (c) Generated summary: an automatically generated summary from all reviews, (d) Comment context: highlights the context of each argument within a review from the related paper.}
        \label{fig:modified}
    \end{figure*}
\subsection{Interview Methodology}
We conducted 30 to 45-minute remote semi-structured interviews via Zoom, which were recorded with participants' permission and transcribed. The interviews were structured into two main parts. The first half focused on understanding each participant’s experience as an AC. The second half discussed how ACs envisioned using AI-assistive tools. The interview transcript is provided in the Sect.~\ref{sec:script}.

\subsubsection{AC Experience and Workflow:} Interviews began by asking each participant about their experience as an AC for ICLR. Next, we asked ACs how they viewed their role and the specific tasks in their workflow. These open-ended questions allowed participants to define their responsibilities on their own terms and to surface the challenges they encountered at each stage of the process. They also discussed their strategies and shared their perceptions of what is effective and less effective when managing different reviewing cases.

\subsubsection{Design Probe:} In the second part of the interview, participants were asked to describe their view of a potential AI-assistive tool and the capabilities they envisioned for it. To elicit design considerations, we employed a design probe using four wireframes (see Figure~\ref{fig:modified}) and presented them to participants as static slides in an electronic slideshow. These wireframes were not intended as final system designs but rather to spark discussion with participants~\cite{yudhanto_designing_2022,zaidi_sociotechnical_2025,biermann_tool_2022}. 
We adopted this approach over interactive evaluation to focus on conceptual possibilities and prevent participants from being distracted by technical issues such as latency or interface details.

The order in which each wireframe was shown to participants was randomized to maximize the breadth of design discussions and mitigate order effects. For each wireframe, the interviewer explained the design concept and presented corresponding slides illustrating the features of an AI-assistive tool. After each design concept, the interviewer used probing questions to deepen understanding of how ACs perceived the tool's advantages and disadvantages and how it fits within their workflows. Participants then ranked the designs according to perceived usefulness. The interviewer then asked follow-up questions about design considerations for assistive tools for ACs.

\subsection{Creating the Design Probe}
\label{sec:probe}
We constructed our design probe to present four major conceptual AI-assistive tools based on common challenges faced by ACs in prior work and grounded in AI and HCI literature for peer review. We then conducted brainstorming sessions (approx. 4 hours in total) to sketch the initial wireframes. During brainstorming, our research questions served as a starting point and an anchor. While we incorporated several assistance methods, we did not include those that fully automate AC decision-making, as we target tools that augment rather than replace human judgment.  
The initial sketches were iteratively refined into the final versions presented to participants (see Sect.~\ref{sec:key} for a summary of key changes and an illustration of the sketches before and after our iterations). The wireframes are discussed below:

\subsubsection{Argument labels (Fig.~\ref{fig:argument})} 
Several systems have supported decision-making by visually highlighting task-relevant features and annotating them with descriptive labels via color-coded tags~\cite{bansal_does_2021,fok_scim_2023,fogliato_who_2022}. This design feature is particularly valuable when users must process long documents, as it enables quick identification of relevant sections without exhaustive reading. Empirical evidence supports these benefits: in one study, it was found that color-coded tags helped participants locate relevant information faster and maintain focus\cite{fok_scim_2023}. In our context, ACs navigate long reviews at different stages, first to identify topics discussed during rebuttal, and later to synthesize reviewers' arguments when drafting their meta-review~\cite{wang_sentiment_2018,ghosal_deepsentipeer_2019,hua_argument_2019}. Thus, we developed a wireframe that color‑codes each review argument by the aspect discussed (e.g., method) and marks its associated sentiment with a positive (+) or negative (–) sign. The wireframe offers an alternative view (shown in the Sect.~\ref{sec:key}, Fig. 5-C) that clusters all arguments related to the same aspect, with each argument accompanied by a reviewer's identification.

\subsubsection{Conformity to standards (Fig.~\ref{fig:standards})} 
 A substantial part of the AC’s decision-making process involves weighing reviewers’ arguments and flagging potential issues. To address these challenges, AI methods have been developed to evaluate review quality~\cite{arous_peer_2021,flach_novel_2010,tan_least_2021}. Similar AI-generated predictions have been used to assist decision-makers in diverse domains: for instance, systems have shown predicted biometrics to physicians when prescribing medications~\cite{burgess_healthcare_2023}, toxicity scores to content moderators~\cite{chandrasekharan_crossmod_2019}, risk assessments to judges making bail decisions~\cite{grgic-hlaca_human_2019}. Therefore, we design a wireframe that shows a score reflecting a review's conformity to conference standards that covers three dimensions: 1. \textbf{Clarity} evaluates if a review is well-structured and includes supporting arguments for the recommended score; 2. \textbf{Consistency} checks alignment between the recommended score and the review content; and 3. \textbf{Tone} evaluates the politeness of a review.

\subsubsection{Generated summary (Fig.~\ref{fig:generated})}
Summaries have been shown to reduce cognitive load and allow decision-makers to complete tasks more efficiently~\cite{han_ascleai_2024}. They have been applied in domains such as legal case analysis~\cite{norkute_towards_2021}, research ideation through literature-based summaries~\cite{pu_ideasynth_2025}, and clinical review by condensing patient–doctor dialogues~\cite{han_ascleai_2024}. Building on this principle, various LLM-based methods have been developed for meta-review generation~\cite{li_towards_2023,li_sentiment_2024,zeng_scientific_2025}. In our design, we opted for extractive summarization based on early brainstorming discussions with team members who had AC experience. They emphasized the risks of hallucinations associated with abstractive summarization and preferred to retain the original wording of reviewers’ arguments. Therefore, our wireframe proposes a method that clusters reviewers’ arguments to highlight areas of agreement and disagreement, while separately compiling additional comments that appear when a reviewer raises a unique point.
 
\subsubsection{Comment context (Fig.~\ref{fig:context})}
Various reading interfaces aim to enhance decision-making by presenting contextual evidence that users would otherwise need to search for or extract through extensive reading. For example, some tools highlight relevant patient histories in clinical settings~\cite{cai_hello_2019}, while others provide in-context definitions of technical terms in academic papers~\cite{head_augmenting_2021}. Inspired by these designs, we illustrated a wireframe that enables navigation of reviewer–author discussions by highlighting the paper's context for each argument and displaying the corresponding author response.

\subsection{Analysis}
We employed reflexive thematic analysis to analyze the interview data~\cite{braun_thematic_2021}. Our process involved several iterative stages. First, three of the authors read the transcripts to familiarize themselves with the data. They then inductively developed codes through open coding, identifying segments of data potentially relevant to the research questions and assigning descriptive labels. Second, they iteratively refined existing codes by comparing them to the interview transcripts. Next, the coders clustered related codes into preliminary themes that captured recurring patterns or concepts. To strengthen the analysis, all authors engaged in multiple synchronous meetings to discuss the themes that were actively created and iteratively refined them, ensuring that each one was ``clearly demarcated and built around a strong core concept''~\cite{braun_thematic_2021}. Ultimately, we created 162 codes and 6 main themes, such as: `a challenging dynamic between passive reviewers and persistent authors, and ACs’ frustration with chasing responses'; and `the scalability challenges in peer review with ACs' cognitive overload'. We continually assessed the alignment of the developed themes with our research questions. We organized the findings according to our research questions. For RQ1 (challenges and strategies), we iterated until we had a set of distinct themes that captured participants' experiences. For RQ2 (feedback on the design probe), we grouped responses into advantages, concerns, and desired design features, refining these categories until we obtained a representative set.

\section{Findings}

\subsection{Existing Challenges (RQ1a)}

\subsubsection{Navigating Tensions in Managing Reviewers and Authors}
\label{sec:dynamic}
ACs are tasked with managing a challenging dynamic between reviewers and authors: reviewers contribute their time and expertise without direct incentives, while authors are eager for their papers to be accepted.

\noindent\textbf{Reviewers' lack of commitment.}
\label{sec:volunteer}
 Participants admitted that reviewers often lacked commitment during the review process. They frequently submitted their reviews late, or not at all, often without notifying ACs. This lack of commitment was also reflected in the quality of reviews: reviewers sometimes overlooked key aspects of a paper’s contribution or confidently judged it without clear justification. Such shortcomings were evident in short, generic, and poorly justified reviews.
Among the 27 interviewed ACs, 22 reported that reviewers often disengaged from the process after submitting their reviews, despite AC reminders. Their opinions remained unchanged after the rebuttal or even ``ghosted" the authors' responses. One AC admitted:
\begin{quote}
    P1: It's sort of running back and forth digitally between folks, and I think it'd be much nicer if reviewers engaged more without the AC having to mediate.
\end{quote}
Some ACs attributed reviewers' lack of engagement to the limited time reviewers have, as ``everything is as rushed as possible." They also hypothesize that these issues stem from a lack of incentives, as reviewers felt obligated to give back to the community but did not receive tangible benefits.

\noindent\textbf{Authors' persistence.}
ACs noted that authors needed to push their papers for publication, as it was required to advance their careers. This created an asymmetric situation in which authors have strong incentives to engage in extensive discussion, whereas reviewers do not. Additionally, ACs observed a range of strategies employed by authors to persuade reviewers to reconsider their scores or to convince ACs that reviewers had misunderstood their work.

\begin{quote}
    P19: You [as an author] are very polite when writing to the reviewers. But then you're mean when writing to the AC, because your goal is either to convince the reviewer or to discredit the reviewer with the AC.
\end{quote}
\noindent\textbf{Herding cats.} 
Managing this challenging dynamic between reviewers and authors came down to the ACs. They reported various strategies to ``chase" reviewers, including sending reminders (\textit{‘it’s due in two days!’}, \textit{‘it’s due tomorrow!’}) and contacting reviewers through alternate channels such as Slack or their personal email. Participants found this task frustrating, noting that it did not require their level of seniority and left them feeling undervalued.
\begin{quote}
P13: You don't need actually a lot of subject-specific knowledge, and it is a bit of a waste of time. 
\end{quote}

\subsubsection{Scalability Challenges in Peer Review}
\label{sec:snowball}
The rapid increase in submissions poses challenges for peer review.

\noindent\textbf{Scarcity of expert reviewers.}
\label{sec:exprt}
The disparity between the number of expert reviewers and the number of submissions frustrated ACs, as they received fewer qualified reviews than needed for a fair evaluation.
To address this issue, various conferences recruited junior researchers as reviewers, which ACs viewed as ``suboptimal". For instance, \codecount{5} ACs found that junior reviewers tend to focus excessively on novelty and often hold extreme opinions: they either judged harshly on minor details, expecting groundbreaking contributions, or were overly positive, overlooking fatal flaws or existing work.

\noindent\textbf{Cognitive overload.}
\label{sec:heavy_load}
Participants described the difficulty of managing a large volume of submissions,  sometimes being assigned over 20 papers for a conference. This heavy workload, combined with their busy schedules and the varying quality of reviews, made it difficult for them to track each paper. 
They highlighted the challenges of navigating the extensive information required for a fair and informed decision. Among them, \codecount{10} ACs aimed to collect as much information as possible and ensured that everyone involved had a fair opportunity to articulate their views. Gathering all this information involved facilitating productive discussions and navigating numerous detailed technical responses across different interfaces. Handling this information overload, particularly in conferences with lengthy discussions, was overwhelming.
\begin{quote}
    P7: It’s sometimes kind of a cognitive overload to see what's going on in the discussion.
\end{quote}

\subsection{ACs' Strategies without Tool Support (RQ1b)}

\subsubsection{Nudging Reviewers}
ACs admitted that their most frustrating task was coordinating between reviewers and authors. They primarily sent reminders and prompted reviewers to update their reviews and participate in discussions. 
\begin{quote}
    P6: On the one hand, kind of bureaucratically, there's like managing this, and part of it is like all the herding cats and making sure that the reviewers do their job.
\end{quote} 
In these communications, many ACs used template messages, either provided by the conference or developed from their prior experience, to save time. Among our participants, \codecount{16} ACs aimed to resolve disagreements and reach a consensus by encouraging reviewers with strongly divergent views or weaker arguments to read the other reviews, sometimes by explicitly prompting them. They also noted that the discussion would sometimes veer off course, necessitating their intervention to maintain a professional tone.

\subsubsection{Varying Strategies in Interpreting Reviews}
In cases of disagreement among reviewers, we found that ACs used different strategies to weigh reviewers' arguments. For instance, \codecount{5} ACs relied on the opinions of reviewers whose expertise closely matched the paper's topic. Another \codecount{18} ACs relied on long and detailed reviews that provide concrete evidence for their claims and actively participate in the discussion. \codecount{3} ACs used authors' responses to assess the quality of reviews. Some ACs noted that some reviews reflected personal opinions or biases toward specific research areas rather than objective assessments. To address this, ACs described actively filtering facts from opinions and copying comments from reviews into a document or spreadsheet, organizing them by topic or severity.
They also evaluated the extent to which the rebuttal addressed reviewers' concerns.

\subsubsection{Selective Engagement with Submissions and Decisions}
Across the interviews, we observed that ACs vary widely in how much they engage with submissions and in their level of involvement in the outcomes. For instance, \codecount{14} ACs leveraged their seniority and network in the decision-making process. They tended to review papers, advocating for promising ideas that can stimulate engaging discussions at conferences. They also sought to educate reviewers and help them advance in their roles.
Other ACs distinguished their role from that of reviewers and were reluctant to read papers or provide their own opinions on submissions. Instead, they attempted to be impartial, listening to all sides and making a decision based on consensus and what they found most convincing. 
Several ACs admitted struggling with the extent to which they need to remain impartial rather than injecting their own opinions, where \codecount{5} ACs explicitly expressed concern about whether it was appropriate to overrule reviewers, especially when they disagreed with reviewers' consensus.
\begin{quote}
 P17: There are times when I think all the reviewers are wrong, and I disagree with them. Is it still my job to, like, maybe try to overturn the reviewers, or just ignore them and do what I want?
\end{quote}
These variations were also reflected in the reported time spent on AC duties: some reported spending about 10 minutes per paper, while others estimated 45 minutes to several hours, with one AC dedicating a full week of work to these duties. Some ACs were aware of these variations and saw potential in using tools to make the process more effective.
\begin{quote}
    P13: Maybe someone is hands-off, and they just don't really do anything until the very end, or someone is hands-on, and maybe they're spending a disproportionately large amount of time on the task [...] tools, slash training, and all that, presumably could play some role in making the process a bit more effective.
\end{quote}

\begin{figure*}
    \centering
    \includegraphics{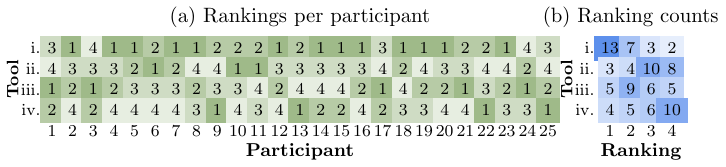}
    \caption{
     Stated preference rankings for the wireframes presented (i. argument labels; ii. conformity to standards; iii. generated summary;  iv. comment context) for individual participants (a) and aggregated across participants (b).
    }
    \label{fig:rankings}
\end{figure*}
\subsection{Perceptions of AI Tools and Ethical Concerns Among ACs (RQ2)}
While reviewing the wireframe prototypes (Fig.~\ref{fig:modified}), ACs discussed potential ways in which AI‑assistive tools could support them. During these discussions, many ACs envisioned tools and workflows that extended beyond the presented designs. Alongside these opportunities, ACs expressed varied preferences and raised ethical concerns about deploying such tools. We therefore report both wireframe‑elicited responses and emergent design ideas, as well as participants' concerns. We report the ACs' ranking of the wireframes by usefulness in Fig.~\ref{fig:rankings}.

\subsubsection{Wireframe‑Elicited Visions}
\label{sec:wireframe}
We first present design concepts elicited through the four wireframe prototypes:\\
\noindent\textbf{Argument labels.} When shown this wireframe (Fig.~\ref{fig:argument}), the majority of ACs responded positively to the visual representation of reviewers’ arguments. Several noted that the interface appeared to be ``replacing what an AC would do [...] barring the actual decision". ACs appreciated that the interface streamlined the review process by enabling quick skimming and providing a structured view of reviewers’ concerns. They found that such an interface could assist in the decision-making phase by identifying common weaknesses in a submission, thereby making it more efficient to form their opinion. Additionally, they recognized its potential to simplify writing meta-reviews, as they would use the highlighted arguments from the reviews as evidence for their decision. At the same time, ACs expressed concerns about \emph{over‑reliance} on the tool. As experts in AI, they cautioned that errors in argument classification could lead ACs to misinterpret reviewers’ intent, potentially resulting in flawed judgments. Moreover, several ACs worried that presenting arguments in a structured, summarized form might reduce the decision-making process to counting pros and cons rather than thoroughly assessing the reviews. 
\begin{quote}
    P18: I'm not sure if the algorithm has grouped one sentence to a wrong class which leads, to a wrong direction, then lead me to a wrong decision. Who will be responsible for such a mistake?
\end{quote}

    ACs’ design feedback emphasized the importance of visualizing reviews ``as‑is," rather than interrupting or restructuring their narrative flow. Several participants expressed concern that the alternative interface shown in the Sect.~\ref{sec:key}, Fig.5-C, which segmented and reordered review content, could increase the risk of misinterpreting reviewers’ arguments. Participants also suggested enabling both ACs and reviewers to modify the categorizations of reviewers’ arguments. They emphasized that such edits were necessary to ensure that the overall process remained trustworthy: 
\begin{quote}
P09: I would prefer that any automated judgment come as early as possible in the process, so that there’s as many eyeballs as possible—human eyeballs.
\end{quote}

\noindent\textbf{Conformity to standards.}
\codecount{8} ACs expressed enthusiasm for the conformity to standards interface (Fig.~\ref{fig:standards}), which assigns a score to review quality based on multiple criteria. In terms of workflow integration, ACs noted that the interface could help them proactively identify reviews with toxic language or low-quality content. Others described using the scores to identify submissions that may require additional review, particularly when existing reviews scored poorly. By surfacing such issues early, the interface was seen as a way to intervene before reviews progressed to later stages of the process. Beyond the writing phase, ACs described using the aggregated conformity scores to ``upweight or downweight" individual reviews, treating the signal as contextual information when synthesizing reviews. However, ACs were concerned that scoring inaccuracies could lead to unfair outcomes by downweighting certain reviews. Others raised concerns about the potential for \emph{bias propagation}, particularly if existing reviewer biases were amplified through automated scoring mechanisms.  
   \begin{quote}
        P1: Especially as somebody who does machine learning, I wanna know where these numbers came from. 
    \end{quote}
Drawing on their expertise in machine learning, participants described how such a tool could benefit from \emph{interpretable} models that allow ACs to understand the generated evaluations, rather than treating scores as opaque judgments. In addition, they suggested adding a \emph{personalization} feature that enables the tool to adapt to individual ACs’ preferences. For instance, several ACs suggested incorporating additional signals, such as review length, technical soundness, or reviewers’ expertise. ACs also suggested adding a feature to enable them to calibrate the system based on their own assessments of review quality. For example, some envisioned adjusting weights in response to how closely the system’s evaluations aligned with their personal judgments over time.

\noindent\textbf{Generated Summary.}
\codecount{12} ACs considered the generated summary a useful tool that could substantially save them time if it faithfully conveys reviewers' concerns. They found that they could use it to have a first glance at all reviews, as a starting point to writing the meta-review, or to verify that they did not miss any comments. Despite its potential to save time, ACs were concerned about the possibility of it becoming an ``automatic AC." They believed that the meta-review should go beyond summarizing reviews and instead provide feedback to authors. \codecount{6} ACs were concerned with the risk of hallucination, where the generated content might deviate from the original reviews and fail to convey the significance of each argument. Moreover, many expressed apprehension that relying too heavily on such a tool could foster ``laziness," with ACs relying on it exclusively for their meta-reviews, potentially resulting in unfair decisions. 

To mitigate these risks, participants emphasized the importance of ensuring that ACs remain the central decision-makers, with a tool that provides assistance, akin to \emph{human-in-the-loop} systems. They stressed that automation should assist sense-making without replacing expert judgment. Participants recommended interface designs that require ACs to actively edit or revise the generated summary rather than allowing direct copy‑and‑paste, to encourage deliberation and accountability. Additionally, they suggested enabling reviewers to validate the generated summary to ensure it accurately reflects their arguments.
\begin{quote}
    P14: 
    Making sure you force them [ACs] to be an active component of the loop [human in the loop].
\end{quote}

\noindent\textbf{Comment context.}
\codecount{12} ACs identified the \emph{comment context} interface as a valuable time-saving tool, particularly for navigating lengthy review discussions. They appreciated that it reduced the need to manually search through extensive exchanges to locate the context surrounding reviewers’ concerns. Additionally, they noted its usefulness in monitoring the discussion between authors and reviewers. At the same time, ACs cautioned that this solution could not be ``one‑size‑fits‑all." Several emphasized that the interface would need to integrate with their existing reading practices, such as PDF viewers or printed submissions, rather than imposing a separate workflow. Others raised concerns about the reliability of AI‑based contextualization, noting that inaccuracies could lead the system to omit relevant text or link comments to incorrect contexts.

In terms of design features, participants described how such a tool could help them keep up with ongoing discussions by explicitly highlighting arguments that require AC attention. Rather than scanning entire threads, ACs envisioned being proactively directed to comments or exchanges that raised unresolved concerns, requested clarification, or introduced new substantive points through the interface. Others suggested augmenting this functionality by visually marking ``hot spots" within a submission, such as areas where multiple reviewers raised concerns. By surfacing these regions, the tool could help ACs focus their reading on contested areas, supporting more efficient sense-making.

\subsubsection{Emergent Design Ideas Beyond the Wireframes}
\label{sec:emergent}
In addition to responding to the wireframe prototypes, ACs articulated new design ideas. These emergent concepts arose organically through reflection and discussion.

\noindent\textbf{Assisting in reviewers' communication.}
When discussing the potential of AI-assistive tools, seven ACs identified reviewer communication as one of the most frustrating and time‑consuming tasks that can be automated. In particular, they described the need to repeatedly follow up with reviewers to confirm responsiveness as largely ``procedural" work that does not require ACs’ expertise and could be ``easily automated." ACs envisioned tools that support reviewer communication throughout multiple phases of the review process. During the initial review period, they suggested lightweight responsiveness signals that allow reviewers to indicate continued commitment, such as a simple acknowledgment: ``I’m going to do this, but I just haven’t done it yet", which one AC noted could substantially reduce anxiety around missing reviews. During the rebuttal phase, participants proposed automated reminders prompting reviewers to read and respond to authors’ rebuttals. For example, one AC described a web‑based form with a checkbox that would allow reviewers to indicate whether the rebuttal changed their position. One AC framed such a system humorously, revealing underlying frustration with the process:
\begin{quote}
    P06: A robot that could hunt down my reviewers and beat them or a way more annoying alert system to get people to do the work they’re supposed to do. 
\end{quote}

\noindent\textbf{Assisting reviewers and fostering accountability.}
Given the need to recruit junior reviewers, \codecount{10} ACs suggested training programs or tooling to scaffold review writing, arguing that supporting reviewers earlier in the process could reduce the burden on ACs later. 
At the same time, participants highlighted challenges stemming from the voluntary nature of peer review, which led to inconsistent commitment among reviewers. To address these challenges, \codecount{6} ACs suggested leveraging conference records or a tool such as the conformity to standards (Fig.~\ref{fig:standards}) to track reviewers’ reliability and inform future reviewer selection.  

\section{Discussion}
\subsection{Beyond One‑Size‑Fits‑All (RQ1)}
Our findings reveal that ACs vary widely in how much they engage with submissions and in their level of involvement in the outcomes. This variability resonates with prior work on peer review, which frames it as a constellation of practices encompassing consultative, decision-making, administrative, and moderating activities~\cite{reinhart_peer-review_2024}. These activities are often led or supervised by ACs~\cite{roth_editorial_2006,hall_journal_2015,galipeau_scoping_2016,dandrea_can_2017}. However, our study extends this literature by demonstrating that AC engagement with these practices is not uniform but somewhat situational, with some adopting a hands-on approach while others are more hands-off. While previous research has examined reviewer engagement and the challenges they face~\cite{lee_uncovering_2020,wilson_how_2022}, our work is the first to foreground the nuanced challenges ACs encounter in navigating these activities.

\subsubsection{Design Implications: Develop Support Tools that Offer Personalized Assistance}
Rather than prescribing a single ``ideal'' practice, future systems should acknowledge and legitimize the variability in ACs' practices while offering strategies to manage tensions, such as balancing advocacy with impartiality. This perspective also helps explain why many ACs emphasized personalization features and why preferences varied across wireframes. Consequently, we argue that \textit{there is no one-size-fits-all solution for assisting ACs}. Individual differences, such as prior experience, familiarity with the submission, and the level of reviewer engagement, likely shape ACs’ needs. While previous AI tools for ACs have primarily focused on scaffolding novices~\cite{sun_metawriter_2024}, our findings point to an opportunity for personalized, context-aware assistance. Findings from Sect.~\ref{sec:wireframe} show that ACs’ preferences for particular wireframes were often grounded in how well these designs aligned with their existing workflows. For instance, the argument-labeling interface was positively received because it supports the structuring work ACs already do, without replacing their final judgment. Accordingly, future designs should account for ACs’ established work practices and seek to integrate AI support in ways that complement, rather than disrupt, these workflows. This need for alignment was also evident in discussions of the comment-context interface, where ACs emphasized the importance of integrating such features into their existing reading practices. This observation is consistent with other studies on AI assistance~\cite{tankelevitch_metacognitive_2024,zhang_rethinking_2024,jacobs_designing_2021}, which show that misalignment between AI assistance and users’ decision-making processes can lead to ineffective use or even abandonment of the system. 

Drawing on findings from the different wireframes, especially `conformity to standard', we argue that AI assistance for ACs should strike a balance between being simple yet informative, allowing ACs to get an overview at a glance while also providing explanations of how the results were generated on demand. This finding resonates with explainable AI assistance systems used in clinical settings, where explanations should be selective and mutable~\cite{corti_it_2024,bertrand_selective_2023}, allowing decision-makers to tweak the granularity of the information.

The discussion on the summarization tool revealed a divide between participants who valued its efficiency gains and those who expressed concern about over-reliance. To balance these perspectives, it is critical to ensure ACs engage critically with AI-generated content. This could be by restricting access to AI-generated content until users have formed their initial judgments, as it was shown to promote active engagement~\cite{fogliato_who_2022} or through personalization mechanisms by enabling users to toggle systems' features and adjust their output to their preferences~\cite{zhang_concepteva_2023}. All these potential personalization features should transparently communicate what historical data is used and offer flexible controls that allow ACs to adjust or limit data usage according to their privacy preferences~\cite{meng_personalized_2018,rezk_way_2023,choi_human-centered_2026}.

\subsection{Facilitating Reviewer Coordination (RQ1)}
Our findings reveal that ACs face persistent coordination challenges, including chasing reviewers for timely responses, managing uneven participation, and navigating disagreements to reach consensus. These dynamics suggest that current communication channels are insufficient for fostering productive discussions and timely decision-making. As discussed in Sect.~\ref{sec:emergent}, ACs found this coordination task the most frustrating, noting that it did not require their seniority. Prior research on group decision-making highlights the importance of structured facilitation, such as the ``diamond of participation'' framework~\cite{kaner_facilitators_2014}, which encourages divergence of ideas before converging toward consensus. Similar challenges have been addressed in online deliberation settings, where conversational agents have successfully moderated discussions by balancing contributions across participants, promoting topic coverage, and guiding groups toward consensus~\cite{bagmar_analyzing_2022,kim_bot_2020,kim_moderator_2021}. These strategies have been shown to enhance deliberative quality and align group consensus with individual opinions, leading to more authentic outcomes.

\subsubsection{Design Implications: Develop Conversational Agents to Act as Moderators}
Building on these insights, we see an opportunity to explore conversational assistant moderators for peer review. Such an assistant could help ACs by structuring reviewer discussions, encouraging participation from less active reviewers, and preventing dominance by a few voices~\cite{bagmar_analyzing_2022,kim_bot_2020}. By offloading coordination tasks to the assistant, ACs could focus on higher-level judgment and decision-making, aligning their efforts with their expertise and reducing feelings of frustration and undervaluation (expressed in Sect.~\ref{sec:dynamic}). Beyond facilitation, the assistant could visualize discussion stages, highlight key points of agreement and disagreement, and recommend contextually appropriate moderator messages~\cite{lee_solutionchat_2020}, as suggested by ACs in Sect.~\ref{sec:emergent}. However, deploying such an agent requires caution. Prior work shows that errors or poorly timed interventions by conversational agents can significantly undermine participants’ willingness to engage~\cite{do_err_2023,kuang_enhancing_2024}. For example, if a conversational agent incorrectly flags a reviewer as unresponsive, reviewers may feel undervalued, leading to disengagement. Conversely, failing to identify unresponsive reviewers may incentivize them to game the system and avoid detection. Other repair strategies could provide explanations, for example, by notifying reviewers when an author has responded and a specified amount of time has passed without a reviewer reply~\cite{ashktorab_resilient_2019}.

\subsection{Tensions Underlying Informed Caution Toward AI Support (RQ2)}
Our findings reveal a persistent cautiousness among ACs toward AI-based support tools. Our participants recognized the potential of AI tools to streamline decision-making, while cautioning that model inaccuracies could undermine the fairness of final decisions. Across participants, there was a strong consensus that human oversight must remain central to peer review. This cautious behavior may be explained by two phenomena documented in prior research. First, decision-makers with higher AI literacy tend to exhibit greater skepticism toward AI recommendations compared to those with lower familiarity~\cite{jacobs_how_2021,burgess_healthcare_2023}. Similarly, our participants are AI experts and are acutely aware of risks and unintended consequences, which likely shaped their cautious stance (as stated by P1 in Sect.~\ref{sec:wireframe}). Second, prior work consistently emphasizes the importance of safeguarding human agency: decision-makers are more likely to adopt AI tools when these tools assist rather than replace their judgment~\cite{corti_it_2024,fogliato_who_2022,kawakami_improving_2022,ma_who_2023,salimzadeh_missing_2023}. Our findings reinforce this principle in the context of peer review, highlighting the need for AI systems that augment rather than automate AC decision-making.

\subsubsection{Design Implications: Implement Human-Centered AI Principles that Safeguard Human Agency}
Participants’ caution toward AI and their emphasis on human agency have important implications for the design of support tools for ACs. Rather than automating or mimicking the AC role, future systems should focus on augmenting human judgment while maintaining transparency and control. This means ensuring that AI outputs are interpretable and traceable to original sources. This aligns with prior research showing that providing explanations fosters trust and accountability in AI-assisted decision-making~\cite{fogliato_who_2022,bucinca_trust_2021,veale_fairness_2018,cobbe_reviewable_2021,holstein_toward_2023}. For example, extractive summaries that preserve reviewers’ arguments should be favored over abstractive summaries that risk introducing distortions. Unlike authors, who appreciated positively framed summaries to help them accept criticism~\cite{yang_understanding_2025}, ACs prefer avoiding distortions introduced by abstractive summaries. Support tools should also help ACs identify overlooked aspects of reviews and discussions. For example, comparing meta-reviews with individual reviews could highlight inconsistencies or gaps that may otherwise go unnoticed.
This resonates with recent work on human–AI collaboration in peer review, where AI was seen as useful for broadening perspectives and reducing reviewers' blind spots~\cite {chen_envisioning_2025}. Most importantly, future designs must maintain ACs’ active engagement and ensure they remain the final decision-makers to prevent over-reliance, for example, by carefully timing when assistance is introduced~\cite{fogliato_who_2022}, by highlighting key arguments rather than abstracting content in summaries~\cite{solomon_customization_2014}, and surfacing relevant passages from reviewer–author discussions to facilitate conflict mediation~\cite{de-arteaga_case_2020}.

\subsection{Venue‑Specific Dynamics and Broader Implications}
While participants offered valuable insights, they do not represent the full diversity of peer‑review practices across venues. Accordingly, we differentiate generalizable patterns from those specific to ICLR.

\subsubsection{What likely generalizes.} Many challenges highlighted by ACs, most notably the increasing number of submissions, the scarcity of expert reviewers, and the resulting cognitive load on ACs, are not unique to ICLR. Similar pressures have been documented across conference and journal ecosystems, including AAAI, AIES, and adjacent research communities~\cite{ellison_aaai_2025,AIES26,stefanidi_literature_2023}. In response, ACs described developing personal heuristics and lightweight methods for interpreting and reconciling reviewers’ and authors’ arguments, a strategy that plausibly translates to other venues. Likewise, the wireframe concepts we explored are designed to streamline sense‑making and may help reduce cognitive burden independently of venue‑specific norms. As such, these designs are promising options for broader adaptation.

\subsubsection{What appears venue‑specific.} Several dynamics observed in our study reflect ICLR’s particular review context, including lengthy reviewer–author discussions during rebuttals, the public nature of reviews, and the use of the OpenReview platform. Together, these characteristics can amplify tensions between reviewers and authors and increase the coordination and moderation work required of ACs. These dynamics may be less pronounced in venues where reviewer–author interaction is more limited or absent. Additionally, because our participants are AI experts, their skepticism regarding model reliability, bias, and inaccuracies may be more pronounced than that of ACs in other domains. This expertise likely sharpened critiques of prediction‑driven features and raised expectations around personalization, transparency, and control. 

\subsubsection{Implications for transfer to other domains.} Taken together, these observations suggest that AI support for AC work should be generalizable but adaptable: core assistance for coordination and sense‑making (e.g., argument structuring, summarization) can generalize, while levels of automation and interpretability should be adapted to venue norms, and personalization features adjusted to individual AC preferences. The cautious stance voiced by ICLR ACs can be read as a safety‑first baseline. Even if other venues exhibit more moderate skepticism, the concerns and desired features identified here (e.g., preserving human agency, enabling personalization, and offering explanations) represent conservative defaults likely to be appropriate across domains. Finally, we observed substantial variability in how ACs approach their work. While we expect this variability to extend beyond ICLR, future research is needed to empirically examine how it manifests across different venues.


\section{Conclusion and Future Work}
In this paper, we investigate ACs' challenges, strategies, and perceptions of AI assistance. Our analysis reveals that ACs vary widely in how much they engage with submissions and in their level of involvement in the outcomes. While some ACs adopt a hands-off approach and prefer to remain impartial, others were hands-on, reading submissions and actively advocating for promising ideas. 
Through a design probe with wireframes, we found that they approached AI support cautiously: while they recognized its potential to reduce cognitive overload and make decisions faster, they were concerned about the risks of over-reliance, misinterpretation, and bias amplification. From these insights, we distilled three design considerations: personalized assistance aligned with ACs’ diverse workflows; conversational agents to act as moderators between reviewers and authors; and human-centered AI principles that safeguard human agency. By centering ACs’ lived practices, this study provides an empirical foundation for assistive, accountable AI in peer review. Future research could investigate interactive prototypes that accommodate diverse needs and review stages. Further, ACs' experience using these tools over time should be understood using a longitudinal study.


\section*{Acknowledgments}
We are grateful to the area chairs who participated in this study for their time and insights.
The research was undertaken thanks in part to funding from the Connected Minds Program, supported by the Canada First Research Excellence Fund, Grant \#CFREF-2022-00010, and the Canada Research Chairs Program.
We also acknowledge the support of the Natural Sciences and Engineering Research Council of Canada (NSERC) Discovery Grant \#RGPIN-2026-06542 and \#RGPIN-2026-06657.
This work was also supported in part by the National Science Foundation (NSF) under award number 2244805, as well as a FRQNT Master's (B1) research scholarship, a FRQNT Doctoral (B2X) research scholarship, and a NSERC Post-graduate doctoral (PGS D) scholarship.
Any opinions, findings and conclusions or recommendations expressed in this material are those of the authors and do not necessarily reflect those of the sponsors.

\bibliography{references,websites}
\newpage

\section{The ICLR Reviewing Process}
\label{sec:iclr_process}

\begin{table*}[ht]
\caption{Phases of the review cycle in which area chairs are active. The time period of each phase in ICLR 2021 is shown for illustration.}
\label{tab:iclr_rev}
\begin{tabular}{@{}lll@{}}
\toprule
Phase                       & Tasks                                                                & \makecell{Time period\\(ICLR 2021)}                    \\ \midrule
\multirow{4}{*}{Assignment} & \textbf{ACs} bid for and are assigned papers.                                                & \multirow{4}{*}{\makecell{Oct 2 - Oct 12\\(10 days)}}  \\
                            & Reviewers bid for papers.                                                           &                                            \\
                            & \textbf{ACs} inspect tentative reviewer assignments, and modify them if necessary.          &                                            \\
                            & Reviewers receive review assignments.                                               &                                            \\ \midrule
\multirow{2}{*}{Writing}    & Reviewers write and submit reviews.                                                & \multirow{2}{*}{\makecell{Oct 12 - Nov 10\\(29 days)}} \\
                            & \textbf{ACs} address early issues before reviews are sent to authors.                       &                                            \\ \midrule
\multirow{3}{*}{Discussion} & Authors receive reviews and submit rebuttals.                                       & \multirow{3}{*}{\makecell{Nov 10 - Nov 30\\(20 days)}} \\
                            & Reviewers and authors engage in further discussion.                                 &                                            \\
                            & \textbf{ACs} oversee and manage discussions.                                                &                                            \\ \midrule
\multirow{2}{*}{Decision}   & \textbf{ACs} synthesize information in reviews, rebuttals, discussion, and possibly papers. & \multirow{2}{*}{\makecell{Dec 1 - Dec 11\\(10 days)}}  \\
                            & \textbf{ACs} write metareviews and make an accept/reject recommendation.                    &                                            \\ \bottomrule
\end{tabular}
\end{table*}
ICLR is one of the top-tier machine learning conferences, receiving over 11,000 submissions in 2025. The reviewing process at ICLR spans three to four months, from the paper submission deadline to the decision notification date, as detailed in Table~\ref{tab:iclr_rev}. Assignments of reviewers to papers are based on a combination of paper matching scores derived from reviewer profiles, reviewers’ bids on papers, and recommendations from ACs. The process adheres to a double-blind review policy, maintaining anonymity between authors and reviewers. ICLR adopts a collaborative peer-review process in which authors and reviewers can interact on the OpenReview platform\footnote {\url{https://openreview.net/group?id=ICLR.cc}} during the discussion phase, with all exchanges publicly available. This feature is unique to ICLR, as it is among the few machine learning conferences to date that permit public discussion during the review process. This openness has positioned ICLR as a frequent testbed for AI-based peer-review models, enabling researchers to train and evaluate their systems on real-world peer review metadata.

\section{Interview Script}
\label{sec:script}
To guide our semi-structured interviews with ACs, we developed a script that ensured consistency across sessions while allowing flexibility for follow-up questions. The script was designed to elicit insights into ACs’ workflows, challenges, and perspectives on AI-assisted tools. Each interview began with a brief introduction and consent procedure, followed by two main sections: (1) reflections on AC experiences and decision-making strategies, and (2) feedback on AI assistant features with the design probe. The full interview protocol is provided below.
\subsection{Pre-Interview Setup}
\begin{itemize}
    \item Introduce ourselves and the purpose of the study.
    \item Confirm DocuSign completion (if not already done).
    \item Obtain verbal consent and begin audio/video recording.
    \item Introduce the study goal: \textit{“We aim to understand bottlenecks in the peer review process, identify tasks where conference organizers need support, and explore how computational methods can assist.”}
    \item Mention post-interview survey 
\end{itemize}

\subsection{Part 1: AC Experience and Workflow}
\begin{itemize}
    \item How was your experience as an Area Chair (AC) in ICLR?
    \item Are there parts of ACing you find enjoyable?
    \item In your own words, what are the roles of an AC?
    \item From your experience, what is a typical workflow for an AC?
    \item How many papers were assigned to you, and approximately how long did it take to handle them?
    \item How do you handle conflicting information in reviews?
    \item What were the main bottlenecks in managing the reviews?
    \item Can you describe a difficult ACing experience from the past 12 months?
    \item Did you encounter biased reviews? If so, how did you handle them?
\end{itemize}

\textbf{Optional Questions}
\begin{itemize}
    \item What aspects do you pay most attention to when reading reviews?
    \item Do you recall any cases where the review process significantly changed the final decision?
    \item How has your strategy evolved over time?
\end{itemize}

\subsection{Part 2: AI-Assisted Tools and Design Feedback}
\begin{itemize}
    \item What capabilities would you expect from a tool that simplifies the meta-reviewing process?
    \item Please rank the following tools in order of usefulness:
    \begin{itemize}
        \item A conformity check of a review to conference standards (aspects: clarity, consistency, tone—fairness, offensiveness)
        \item Visual representation of argument structure within reviews and clustering of arguments across reviews.
        \item Automatically generated summary draft highlighting main strengths and weaknesses from reviews.
        \item A visualization of the comment context that highlights the relevant sections from the submission for each argument.
    \end{itemize}
    \item For each tool, ask the following question:
        \begin{itemize}
            \item If useful, how would you integrate it into your workflow?
            \item Are there any risks in using such a tool?
            \item If not useful, why?
        \end{itemize}
    \item Do you prefer a human-in-the-loop system for these tools or fully automated ML?
    \item What are the risks or opportunities of using a conversational agent?
    \item How do you typically reach consensus in difficult cases?
    \item What recommendations would you give to researchers designing tools to assist ACs or improve peer review?
\end{itemize}

\subsection{Closing}
\begin{itemize}
    \item Ask if there’s anything they’d like to discuss that wasn’t covered.
    \item Thank the AC for their time and contributions.
    \item Remind them to complete the survey. 
    \item Ask for feedback on the interview experience.
\end{itemize}

\section{Prototypes}
\label{sec:key}
\begin{figure*}[]
    \centering
    \includegraphics[scale=0.2]{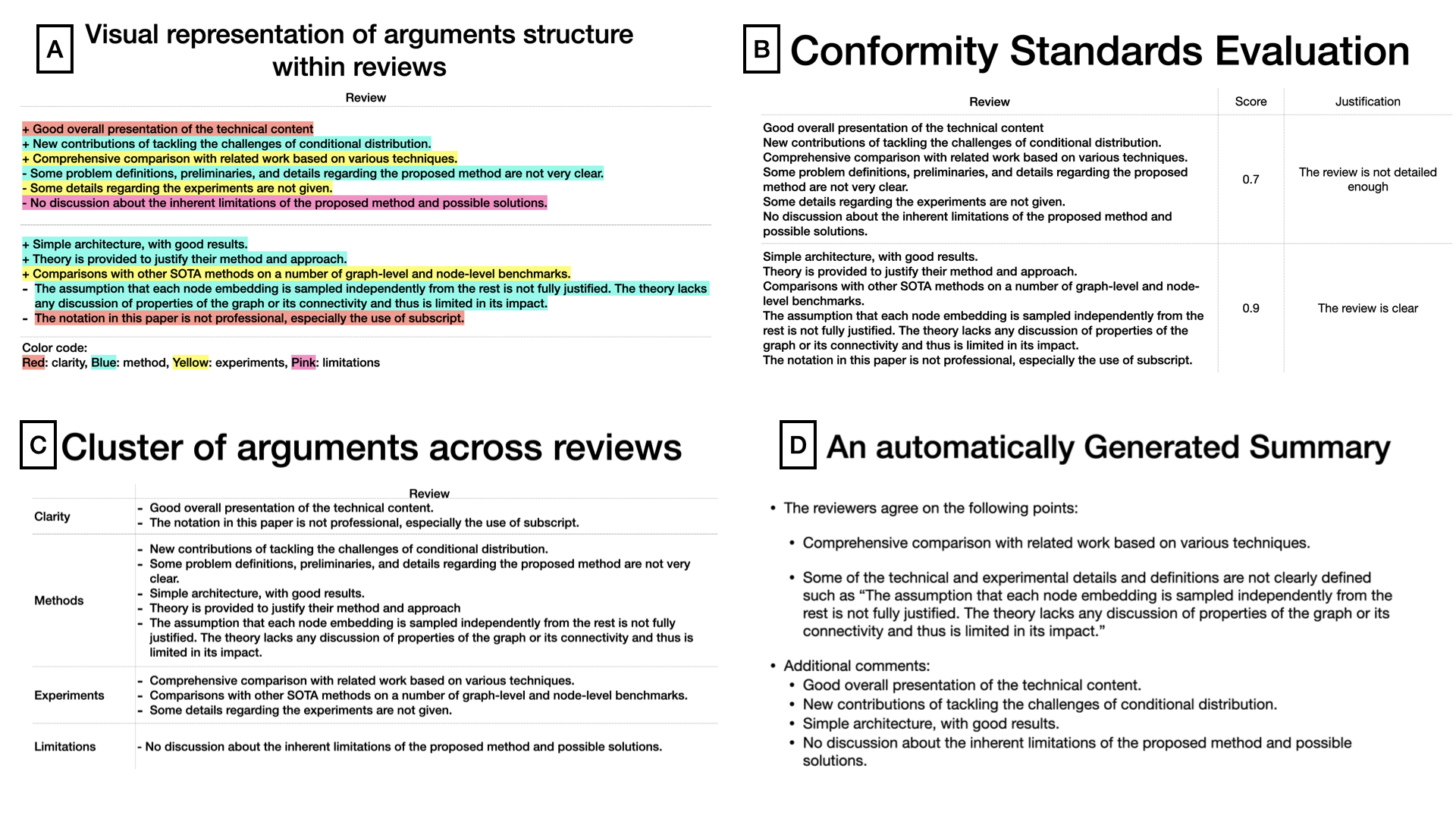}
    \caption{Initial version of the design wireframes: A) Visualization of arguments in each review by highlighting with a color corresponding to a label (clarity, method experiments, limitations). Each argument is preceded by a sign indicating the sentiment. B) Visualization of conformity standards where each review is associated with a score and a brief justification for the score. C) Clusters all arguments belonging to the same topic. D) An automatically generated summary of all reviews.}
    \label{fig:initial}
\end{figure*}

\begin{figure*}[]
    \centering
    \includegraphics[scale=0.2]{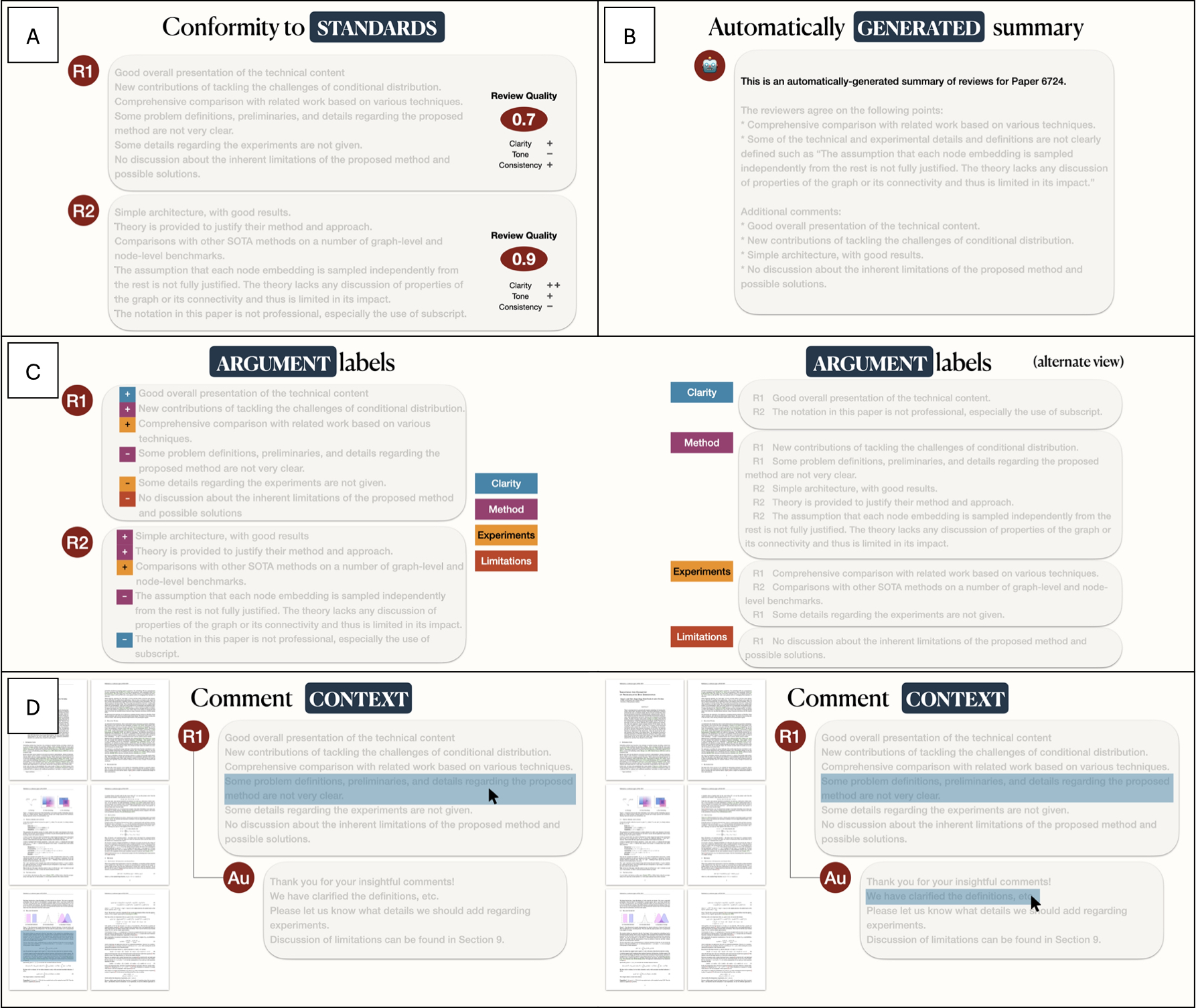}
    \caption{Design wireframes as shown to participants in our interviews.}
    \label{fig:prototypes}
\end{figure*}

When designing our wireframes, we did not include those that fully automate AC decision-making, as we target tools that augment rather than replace human judgment in the meta-review process.  We also merged or discarded design ideas deemed too similar or irrelevant to meta‑review workflows. We ultimately developed four wireframes, each with a corresponding underlying interaction model. These initial designs were iteratively refined through five pilot sessions. These were conducted with two graduate computer science students familiar with interaction design techniques, two team members with AC experience, and pre-interviews with an external AC from the participants' pool. After each pilot interview session, we iteratively refined these designs based on feedback. This allowed us to streamline our designs by resolving confusing elements and avoiding distractions from the intended concept. These wireframes have evolved from their initial version (Fig.~\ref{fig:initial}) to the final version (Fig.~\ref{fig:prototypes}) through the pilot sessions, as follows:
\begin{itemize}
    \item We blurred the review text in the sketches to direct attention away from content details and toward the intended design features.
    \item We merged the visualization of within‑review argument structure (Fig.~\ref{fig:initial}-A) with cross‑review argument clusters (Fig.~\ref{fig:initial}-B), as ACs in our pre-interviews found that they conveyed the same information at different granularities.
    \item In the initial visualization of within-review argument structure (Fig.~\ref{fig:initial}-A), participants in our pre-interviews found the text highlighting distracting and suggested replacing it with color tags placed before the text to indicate topics.
    \item During pre-interviews, ACs emphasized the central role of rebuttal in their decision-making and expressed a need for tools that contextualize the discussion between authors and reviewers. In response, we added a design probe to provide this contextualization (Fig.~\ref{fig:prototypes}-D), which was later validated by the same ACs.
    \item We expanded the explanation of the ``conformity‑to‑standards'' (B) score by specifying the criteria that determine how the score is generated.
\end{itemize}

\end{document}